\documentclass[
aps,
prd,
reprint,
showpacs,
floatfix,
nofootinbib,
superscriptaddress,
preprintnumbers,
letterpaper
]{revtex4-2}

\usepackage{amsmath,bm}
\usepackage{graphicx}
\usepackage{xcolor}
\usepackage{subfig}
\usepackage{hyperref}
\begin{document}

\title{Growth-of-Structure Constraints on the Variable Chaplygin Gas Model}

\author{Rishabh Jain}
\email{rishabhjain1203@gmail.com}
\noaffiliation

\author{Shruti Thakur}
\email{shruti@ststephens.edu}
\affiliation{St. Stephen's College, University of Delhi, India}

\author{Geetanjali Sethi}
\email{getsethi@ststephens.edu}
\affiliation{St. Stephen's College, University of Delhi, India}
\begin{abstract}
A unified description of dark matter and the late-time acceleration of the Universe offers an attractive framework for explaining the dark sector. Among such models, the Variable Chaplygin Gas (VCG) model provides a unified description but faces challenges, particularly in the evolution of cosmological perturbations. We investigate the evolution of linear density perturbations in the VCG model assuming adiabatic perturbations and constrain its parameters using recent $f\sigma_8$ growth measurements within a Markov Chain Monte Carlo (MCMC) framework. Gaussian priors are adopted from our previous analysis of background observations, including Type Ia supernovae (Pantheon), baryon acoustic oscillations (BAO), Hubble parameter $H(z)$ measurements, fast radio bursts (FRB), and gamma-ray bursts (GRB). We find that growth data favour a markedly different region of the parameter space than the background observations. In particular, the growth analysis prefers a substantially larger value of the model parameter $n$ ($n \simeq 2.3$) compared with the background constraint ($n \sim 1$). This tension indicates that parameter values providing an excellent fit to the background expansion fail to simultaneously reproduce the observed growth of cosmic structures. Our results demonstrate that growth-of-structure observations provide a stringent and independent test of unified dark sector models and underscore the importance of combining background and perturbation data when assessing the cosmological viability of the Variable Chaplygin Gas model.
\end{abstract}
\maketitle
\section{Introduction}
Limitations with the standard approach to solve the late time acceleration problem has motivated us to explore other alternative cosmological models. Since modifying the gravity sector of the spacetime (for example f(R) gravity) are highly constrained \cite{Clifton2012,Tsujikawa2008,Thakur2011,Thakur2013}, considering exotic matter as the source of late time accelerations is a possibility which has been widely explored\cite{Copeland2006,Padmanabhan2003}.\\
In this context while the dynamics of background cosmology has been studied extensively, constraining the models using the growth of perturbation has relatively received lesser attention \cite{Linder2005,DESI2024,Euclid2024,Huterer2023,Nguyen2023,Planck2018}. The Chaplygin gas \cite{Kamenshchik2001} model and its variant  the Variable Chaplygin Gas (VCG) \cite{Guo2007,Sethi2024,Bhardwaj2024}, Generalised Chaplygin Gas (GCG) model\cite{Bento2002,Sen2005} are a few examples of exotic source of late time acceleration.  \\ In this paper, we explore the growth of perturbations in the VCG model. Section \ref{sec:2} reviews the standard approach of using VCG in the context of background cosmology. In Section \ref{sec:3}, we present the perturbation equations governing the evolution of cosmic perturbations in the VCG model . Section \ref{sec:4} describes  the observable quantities  to be probed in order to constrain the model parameters of the model. In Section \ref{sec:5}, these results are compared with observational data to establish bounds on the model parameters. Section \ref{sec:6} briefly summarizes the conclusions of this work and discusses their significance. 

\section{Background Cosmology of Variable Chaplygin Gas}
\label{sec:2}
Numerous dark energy models beyond scalar-field descriptions have been proposed in the literature. A broad class of such models is based on barotropic fluids, in which the pressure is expressed as a function of the energy density,
\begin{equation}
p = f(\rho).
\end{equation}
The specific form of the equation of state \(f(\rho)\) determines the properties of the barotropic fluid. A notable example is the Chaplygin gas \cite{Kamenshchik2001}, which is characterized by the equation of state
\begin{equation}
p = -\frac{A}{\rho},
\end{equation}
where \(A\) is a positive constant. \\
Taking the temporal component of the energy-momentum conservation equation,
\begin{equation}
\nabla_{\mu} T^{\mu\nu} = 0,
\end{equation}
one obtains the continuity equation
\begin{equation}
\frac{\partial \rho}{\partial t} + 3H(\rho + p) = 0,
\label{continuityeq}
\end{equation}
where \(H = \dot{a}/a\) is the Hubble parameter and \(\rho\) denotes the total energy density of the Universe.

Using Eq.~(\ref{continuityeq}), the energy density of the Chaplygin gas model can be written as \cite{Sethi2024}
\begin{equation}
\rho_{\rm ch} = \sqrt{A + \frac{B}{a^{6}}},
\label{chaplyginrhoarelation}
\end{equation}
where \(a\) is the scale factor and \(B\)  is an integration constant.

In the early Universe, when \(a \ll 1\), the above expression reduces to \(\rho_{\rm ch} \propto a^{-3}\), indicating a matter-dominated behavior similar to cold dark matter (CDM). In the late-time limit, as the scale factor approaches unity, Eq.~(\ref{chaplyginrhoarelation}) yields \(p = -\rho = \text{constant}\), so that the Chaplygin gas behaves like a cosmological constant. This feature implies that the model can account for the accelerated expansion of the Universe.

Chaplygin gas-based models have continued to attract considerable interest among researchers, owing to the promising features they exhibited in earlier studies \cite{Bhardwaj2024,Bhardwaj2025,Mukherjee2024,Aggarwal2025,AlMamon2026}. However, a known limitation of the standard Chaplygin gas model is that it can lead to oscillations or an exponential growth in the matter power spectrum , which is inconsistent with observational data \cite{Sandvik2004,Bean2003,Fabris2010}. To alleviate this instability, the combined effects of shear and rotation may be taken into account. These effects slow down the collapse relative to the standard spherical collapse scenario \cite{DelPopolo2013}. Motivated by these considerations, the Variable Chaplygin Gas (VCG) model has been proposed as a modification of the original Chaplygin gas model \cite{Guo2007, Sethi2006}.

Several extensions of the Chaplygin gas model have been proposed in the literature, including the Variable Chaplygin Gas (VCG) and the Generalized Chaplygin Gas (GCG) models \cite{Bento2002,Sen2005}. The generalized Chaplygin gas is defined by the equation of state
\begin{equation}
p_{\mathrm{gcg}} = -\frac{A}{\rho_{\mathrm{gcg}}^{\alpha}},
\label{gcg1}
\end{equation}
where \(A>0\) and \(\alpha\) are constants. The requirement \(A>0\) ensures a negative pressure, which is necessary to account for the accelerated expansion of the Universe.

The generalized Chaplygin gas model is characterized by
\begin{equation}
p_{\mathrm{ch}} = -\left(\frac{A}{\rho_{\mathrm{ch}}^{\alpha}}\right),
\label{chaplygineq}
\end{equation}
where \(0 < \alpha \leq 1\). Using Eq.~(\ref{continuityeq}), the corresponding energy density evolves as \cite{Bento2002}
\begin{equation}
\rho_{\mathrm{ch}} = \left(A + \frac{B}{a^{3(1+\alpha)}}\right)^{\frac{1}{1+\alpha}},
\label{gcg}
\end{equation}
where \(a\) is the scale factor and \(B\) is an integration constant.
In the VCG model, the equation of state of the Chaplygin gas is allowed to evolve across different cosmic epochs. The Variable Chaplygin Gas model \cite{Guo2007,Sethi2024} is defined by
\begin{equation}
p_{\mathrm{ch}} = -\frac{A(a)}{\rho_{\mathrm{ch}}},
\end{equation}
where \(A(a) = A_0 a^{-n}\) is a positive function of the cosmological scale factor \(a\), and \(A_0\) and \(n\) are constants. The Variable Chaplygin Gas (VCG) obeys the generalized equation of state \(p=-A(a)/\rho\), where \(A(a)\) depends explicitly on the scale factor. Consequently, the fluid is not strictly barotropic, since its pressure is not solely a function of the energy density. Nevertheless, along the homogeneous cosmological background, where both \(p\) and \(\rho\) evolve as functions of the scale factor, the pressure may be regarded as an effective function of the energy density.

The VCG model is dynamically stable,  and it provides a good fit to several cosmological observations, including CMBR data \cite{Sethi2024}. Using the continuity equation, the energy density of the VCG model is obtained as
\begin{equation}
\rho_{\mathrm{ch}} = \sqrt{\frac{6}{6-n}A_0 a^{-n} + \frac{B}{a^6}},
\end{equation}
where \(B\) is an integration constant.

Redefining
\begin{equation}
B_s \equiv \frac{B}{\frac{6A_0}{6-n}+B} = \frac{B}{\rho_{\mathrm{ch}0}^{\,2}},
\end{equation}
the energy density can be rewritten in the form
\begin{equation}
\rho_{\mathrm{ch}} = \rho_{\mathrm{ch}0}\left(\frac{B_s}{a^{6}}+\frac{1-B_s}{a^{n}}\right)^{1/2}.
\end{equation}
We consider a spatially flat Friedmann--Lema\^{\i}tre--Robertson--Walker (FLRW) Universe. In the Variable Chaplygin Gas (VCG) model, the Hubble parameter can be expressed in dimensionless form as
\begin{equation}
h^2(z) = \left[B_s(1+z)^6 + (1-B_s)(1+z)^n\right]^{1/2},
\end{equation}
where \(B_s\) and \(n\) are the model parameters.

In the present analysis, we assume that the Universe is filled only with the VCG component, while the contributions from radiation and baryonic matter are neglected. The equation-of-state parameter for the VCG model is given by
\begin{equation}
w(z) = -\frac{6-n}{6}\,\frac{1-B_s}{B_s(1+z)^{6-n} + (1-B_s)}.
\end{equation}

At early times (\(z \gg 1\)), the model behaves effectively as pressureless matter, with \(\rho \propto a^{-3}\), whereas at late times it approaches dark-energy-like behavior.


\section{Cosmological Perturbations in VCG}
\label{sec:3}
The evolution of cosmological perturbations is examined within the framework of linear perturbation theory by considering small departures from the homogeneous and isotropic FLRW background. Since the perturbations are assumed to be infinitesimal, only first-order terms are retained. The perturbed line element is therefore given by
\begin{equation}
ds^2 = -(1+2\phi)\,dt^2 + a(t)^2(1-2\psi)\delta_{ij}dx^i dx^j,
\label{pertmetric}
\end{equation}
where \(\phi\) and \(\psi\) are scalar perturbations.

In linear perturbation theory, the most general perturbation of the FLRW spacetime can be decomposed into scalar, vector, and tensor modes. Since these sectors evolve independently at first order, and the growth of large-scale structure is governed solely by scalar perturbations, we restrict our analysis to the scalar sector.

The mathematical form of the perturbed metric depends on the choice of gauge. We adopt the longitudinal (Newtonian) gauge, which provides a particularly convenient description of scalar perturbations. Although the metric representation differs in other gauges, such as the synchronous gauge, physical observables are independent of the gauge choice.

The longitudinal gauge provides a convenient framework for describing scalar perturbations, since the metric perturbations correspond directly to Bardeen's gauge-invariant potentials. In the absence of anisotropic stress, the Einstein equations imply that the two scalar potentials satisfy $\phi=\psi$, so that the scalar perturbations are completely described by a single gravitational potential.
In the present work, we restrict our analysis to adiabatic perturbations, for which
\begin{equation}
\delta p = c_a^2 \, \delta \rho,
\end{equation}
where the adiabatic sound speed is defined as
\begin{equation}
c_a^2 = \frac{\dot{p}}{\dot{\rho}}.
\end{equation}
Accordingly, the physical (rest-frame) sound speed coincides with the adiabatic sound speed,
\begin{equation}
c_s^2 = c_a^2.
\end{equation}
To perturb the Einstein equation, we consider scalar perturbations in the Newtonian gauge. For a perfect fluid, the perturbed energy--momentum tensor is 
\begin{equation}
T^\mu{}_\nu = (\rho+p)u^\mu u_\nu + p\,\delta^\mu{}_\nu,
\end{equation}
with first-order perturbations
\begin{equation}
\begin{aligned}
\delta T^0{}_0 &= -\delta\rho, \\
\delta T^0{}_i &= (\bar\rho+\bar p)\,\delta u_i, \\
\delta T^i{}_0 &= -(\bar\rho+\bar p)\,\delta u^i, \\
\delta T^i{}_j &= \delta p\,\delta^i{}_j.
\end{aligned}
\end{equation}
We define the density contrast as \(\delta \equiv \delta\rho/\bar\rho\) and the sound speed as  \(c_s^2 \equiv \delta p/\delta\rho\). The velocity perturbation is written as
\begin{equation}
\delta u^i = \frac{1}{a}\partial^i v, \qquad \delta u_i = a\,\partial_i v,
\end{equation}
where \(v\) is the scalar velocity potential.

The Fourier-space continuity equation at first order becomes
\begin{equation}
\dot{\delta}
= -(1+w)\left(-\frac{k^2 v}{a} - 3\dot{\phi}\right)
- 3H(c_s^2-w)\delta,
\end{equation}
while the Euler equation is
\begin{equation}
\dot{v} + (1-3c_s^2)Hv + \frac{\phi}{a}
+ \frac{c_s^2}{a(1+w)}\delta = 0.
\end{equation}
Combining the first-order continuity and Euler equations, one obtains a second-order differential equation for the density contrast. 

\begin{multline}
\ddot{\delta}
+ \left(3H(c_s^2-w)+H(2+w-3c_s^2)
-\frac{\dot{w}}{1+w}\right)\dot{\delta} \\
+ 3H\left[(\dot{c}_s^2-\dot{w})
+(c_s^2-w)\left(\frac{\dot{H}}{H}
+H(2+w-3c_s^2)
-\frac{\dot{w}}{1+w}\right)
\right]\delta \\
-3H\dot{\phi}(1+w)(2+w-3c_s^2)
-3(1+w)\ddot{\phi} \\
+c_s^2\frac{k^2}{a^2}\delta
+(1+w)\frac{k^2}{a^2}\phi
= 0
\end{multline}

The sub-horizon regime is defined by
$k/a \gg H$,
which implies
\begin{equation}
\frac{k^2}{a^2} \gg H^2,\quad \dot{H}.
\end{equation}

so that time derivatives of the metric perturbations are suppressed. Under this approximation, the evolution equation reduces to the evolution of the linear density perturbation for a general cosmological fluid is governed by
\begin{align}
\ddot{\delta}
+ \dot{\delta}\left[2H(1-w)-\frac{\dot{w}}{1+w}\right]
+\frac{c_s^2k^2}{a^2}\delta
-(1+w)4\pi G\bar{\rho}_m\,\delta_m
=0,
\end{align}
Using the Poisson equation,
\begin{equation}
\frac{k^2}{a^2}\phi=-4\pi G\bar{\rho}_m\delta_m,
\end{equation}

where $\delta$ denotes the total density contrast, $w$ is the equation-of-state parameter, and $c_s^2$ is the effective sound speed of the fluid.

However, the observational quantity $f\sigma_8$, inferred from galaxy clustering and redshift-space distortion measurements, probes only the growth of the matter density contrast, $\delta_m$, rather than the perturbations of the total cosmic fluid. Therefore, for comparison with observations, it is appropriate to specialize the perturbation equation to the effective clustering dust-like matter component 
for which
\[
w_m=0,\qquad c_s^2=0.
\]
The general perturbation equation, therefore, reduces to the standard evolution equation for the matter density contrast.
\begin{equation}
\ddot{\delta}_m
+2H\dot{\delta}_m
-4\pi G\bar{\rho}_m\delta_m
=0.
\label{mattergrowth}
\end{equation}
In the Variable Chaplygin Gas (VCG) model, which provides a unified description of dark matter and dark energy, we identify the effective matter density $\bar{\rho}_m$ with the matter-like contribution to the VCG energy density. $\bar{\rho}_m$, the VCG behaves effectively as pressureless dust, while the dark-energy-like contribution remains subdominant. We therefore approximate the VCG energy density by its dominant matter-like term,
\begin{equation}
\bar{\rho}_m
=
\rho_{\rm ch,0}
\sqrt{\frac{B_s}{a^6}},
\end{equation}
which is subsequently used to describe the growth of the effective clustering matter component probed by the $f\sigma_8$ observations. 

Here, an overdot denotes differentiation with respect to cosmic time $t$, $H$ is the Hubble parameter. For numerical convenience, the above equation is expressed in terms of redshift $z$, where primes denote derivatives with respect to $z$.
\begin{align}
\delta_m'' 
+ \delta_m'
\left[ \frac{h'}{h}-\frac{1}{1+z}\right]
-  \frac{3\sqrt{B_s}}{2h^2}(1+z)\delta_m
=0 .
\end{align}

The differential equation is solved numerically with the following initial conditions:
\begin{equation}
\delta_m(z = 1100) = 10^{-5}, \qquad 
\delta_m'(z = 1100) = -10^{-8}.
\end{equation}

The solution of this equation provides the evolution of matter density perturbations, which is subsequently used to compute the growth factor and related observables.

\section{Growth Observables}
\label{sec:4}
Density perturbations are observationally probed through the growth of matter fluctuations. The evolution of matter overdensity provides an important and independent test of cosmological models that aim to explain the late-time accelerated expansion of the Universe. 

The growth rate of matter overdensity is defined as:
\begin{equation}
f \equiv \frac{d \ln \delta_m}{d \ln a},
\end{equation}
where $\delta_m$ is the matter density contrast and $a$ is the scale factor.
The growth factor is defined as:
\begin{equation}
D(z) \equiv \frac{\delta_m(z)}{\delta_m(0)},
\end{equation}
both encodes the evolution of matter perturbations. The growth rate and growth factor  $D(z)$ encapsulates the dependence on the underlying cosmological model, and therefore observables derived from it, such as $\sigma_8$, can be used to distinguish between different models.

\subsection{Mass Variance and $\sigma_8$}
\label{sec:A}
Another important quantity characterising density perturbations is the mass variance of the fluctuation field, defined as\cite{BBKS1986,Song2009,Beutler2012, Eisenstein1998, Linder2005,Planck2018}
\begin{equation}
\sigma_R^2 \equiv \left\langle \left( \frac{\delta M}{M} \right)_R^2 \right\rangle 
= \langle \delta_R^2(\mathbf{x}) \rangle,
\end{equation}
where $\sigma_R$ represents the root mean square (RMS) of the density contrast smoothed over a sphere of radius $R$.

A commonly adopted smoothing scale is $R = 8\,h^{-1}\mathrm{Mpc}$, for which the corresponding root-mean-square mass fluctuation is denoted by $\sigma_8$. This scale approximately marks the transition between the linear and non-linear regimes of structure formation, making $\sigma_8$ a standard measure of the present-day amplitude of matter density fluctuations.

The root-mean-square (rms) amplitude of linear matter fluctuations on the comoving scale
$8\,h^{-1}\,\mathrm{Mpc}$ evolves according to the linear growth of matter perturbations,
\begin{equation}
\sigma_8(z)
=
\sigma_{8,0}
\frac{\delta_m(z)}
{\delta_m(0)},
\label{sigma8_evolution}
\end{equation}
where $\sigma_{8,0}\equiv\sigma_8(z=0)$ denotes the present-day normalization and
$\delta_m(z)$ is the linear matter density contrast. Since we solve the perturbation
equation for the Fourier mode corresponding to the comoving radius
$R=8\,h^{-1}\,\mathrm{Mpc}$, we take
\begin{equation}
k_{\sigma_8}=R^{-1}=0.125\,h\,{\rm Mpc}^{-1}.
\end{equation}
Throughout this work, we adopt the Planck 2018 normalization
$\sigma_{8,0}=0.811$ and compare the theoretical predictions for
$f(z)\sigma_8(z)$ with the observational compilation described in
Ref.~\cite{Kazantzidis2018,Kamenshchik2001,BouhmadiLopez2025,Nesseris2017}. The evolution given by
Eq.~(\ref{sigma8_evolution}) follows directly from linear perturbation
theory, where the growth factor is defined as
$D(z)=\delta_m(z)/\delta_m(0)$
\cite{Peebles1980,Linder2005,Song2009, Huterer2023}

In this work, we consider the growth rate of matter perturbations, $f(z)$, together with the root-mean-square mass fluctuation, $\sigma_8(z)$, as the primary observables for constraining the model. These quantities are inferred from measurements of galaxy clustering. Although galaxies are composed of baryonic matter, they trace the underlying matter distribution, which is dominated by cold dark matter (CDM). Consequently, the growth of the dark matter density field is inferred indirectly through observations of galaxies.

Since galaxies are biased tracers of the underlying matter distribution, the observed galaxy clustering differs from the true matter clustering by a galaxy bias factor. This introduces an additional source of uncertainty in the determination of the matter growth rate. However, the combined observable $f\sigma_8$ can be extracted directly from redshift-space distortion measurements without requiring an explicit determination of the galaxy bias \cite{Song2009}. As a result, $f\sigma_8$ provides a robust and nearly model-independent probe of the growth of cosmic structures and has become one of the principal observables for testing cosmological models.

Motivated by this property, we use the combined observable $f\sigma_8$ to constrain the model parameters. The analysis is performed using a dataset comprising 64 measurements in the redshift range \cite{Kazantzidis2018,Kazantzidis2021,lai2026desi} $0.0 < z < 2$, 

The key observable used in this work is:
\begin{equation}
f\sigma_8(z) = f(z)\sigma_8(z).
\end{equation}
We next constrain the model using this data by computing ${\chi}^2$. The ${\chi}^2$ is given by,
\begin{equation}
{\chi}^2 = \Sigma_{i=1}^{64}  \frac{{f\sigma_{8}}_{obs}(z_i)-{f\sigma_{8}}_{th}(z_i,B_s,n)}{\sigma_{f\sigma_8}}.
\end{equation}
One can also consider a likelihood function.
\begin{equation}
L(\theta) = \exp\left({-\frac{{\chi}^2}{2}}\right)
\end{equation}
 Minimum of ${\chi}^2$ or maximum of likelihood corresponds to the best fit values provided by observational data for the model parameters $\theta$.
 

\section{Observational Data and Parameter Estimation Methodology}
\label{sec:5}
We use observational measurements of the growth observable $f\sigma_8(z)$ obtained from redshift-space distortion (RSD) surveys to constrain the model parameters $\theta = \{B_s, n\}$.

The parameter estimation is performed within a Bayesian framework, where the posterior probability distribution is sampled using a Markov Chain Monte Carlo (MCMC) approach. The likelihood function is constructed by comparing theoretical predictions with observational data:
\begin{equation}
\ln \mathcal{L}(\theta) = -\frac{1}{2} \sum_{i=1}^{N}
\left[
\frac{f\sigma_{8}^{\mathrm{obs}}(z_i) - f\sigma_{8}^{\mathrm{th}}(z_i, \theta)}
{\sigma_i}
\right]^2
\end{equation}
where $N = 64$ is the number of data points and $\sigma_i$ represents the observational uncertainties.

To incorporate prior knowledge from background cosmological observations, we adopt Gaussian priors on the model parameters. These priors are motivated by independent constraints obtained from various datasets, including Type Ia supernovae (Pantheon), baryon acoustic oscillations (BAO), Hubble parameter measurements $H(z)$, fast radio bursts (FRB), and gamma-ray bursts (GRB) by  \cite{Sethi2024}. The corresponding constraints are summarized in Table \ref{tab:background_constraints}.

\begin{table}[htbp]
    \centering
    \caption{Prior constraints on the Variable Chaplygin Gas model parameters, $B_s$ and $n$, derived from various background cosmological datasets \cite{Sethi2024}.}
    \begin{tabular*}{\columnwidth}{@{\extracolsep{\fill}}lcc}
        \hline
        \hline
        \textbf{Dataset} & \bm{$B_s$} & \bm{$n$} \\
        \hline
        SNe Ia (Pantheon) & $0.18 \pm 0.10$ & $1.10 \pm 1.15$ \\
        FRB & $0.09 \pm 0.06$ & $0.44 \pm 0.89$ \\
        BAO & $0.16 \pm 0.11$ & $1.06 \pm 1.25$ \\
        $H(z)$ data & $0.05 \pm 0.006$ & $1.46 \pm 0.23$ \\
        GRB & $0.20 \pm 0.11$ & $1.25 \pm 1.17$ \\
        \hline
        \hline
    \end{tabular*}
    \label{tab:background_constraints}
\end{table}

These background constraints are used to construct Gaussian priors on the parameters, thereby incorporating independent information from the expansion history into the growth analysis.

The posterior distribution is given by:
\begin{equation}
\ln P(\theta \mid \mathrm{data}) \propto \ln \mathcal{L}(\theta) + \ln P(\theta),
\end{equation}
which is sampled using an MCMC algorithm.

The best-fit parameters correspond to the maximum of the posterior distribution (equivalently, the minimum of the chi-square), defined as:
\begin{equation}
\chi^2_{\mathrm{min}} = -2 \ln \mathcal{L}(\theta_{\mathrm{best}}).
\end{equation}

The goodness-of-fit is quantified using the reduced chi-square statistic:
\begin{equation}
\chi^2_\nu = \frac{\chi^2_{\mathrm{min}}}{N - p},
\end{equation}
where $p = 2$ is the number of free parameters.

\section{Results and Conclusion}
\label{sec:6}
\begin{figure*}[t]
\centering

\subfloat[Gaussian priors from Pantheon]{%
\includegraphics[width=0.33\textwidth]{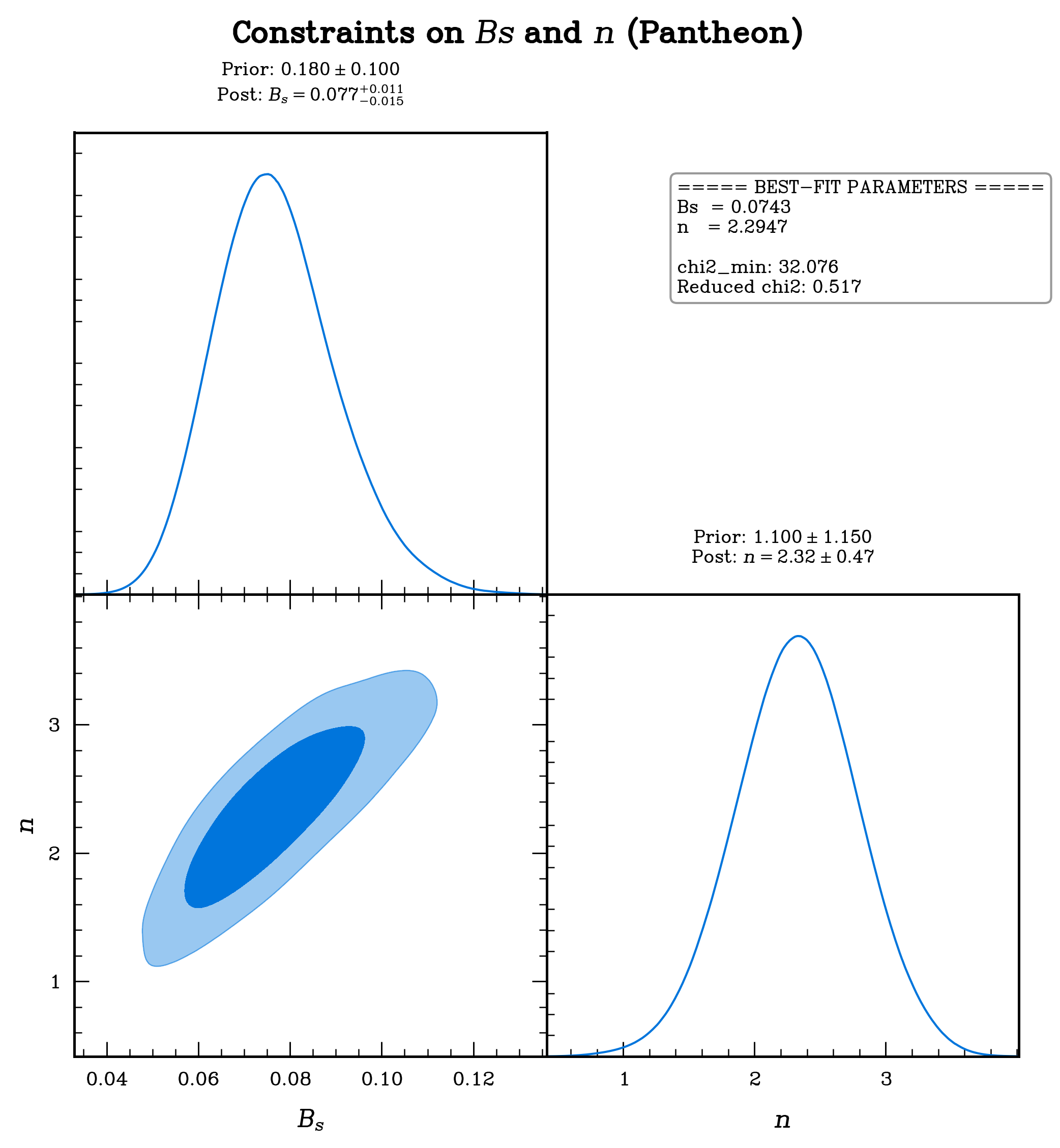}}
\hfill
\subfloat[Gaussian priors from BAO]{%
\includegraphics[width=0.33\textwidth]{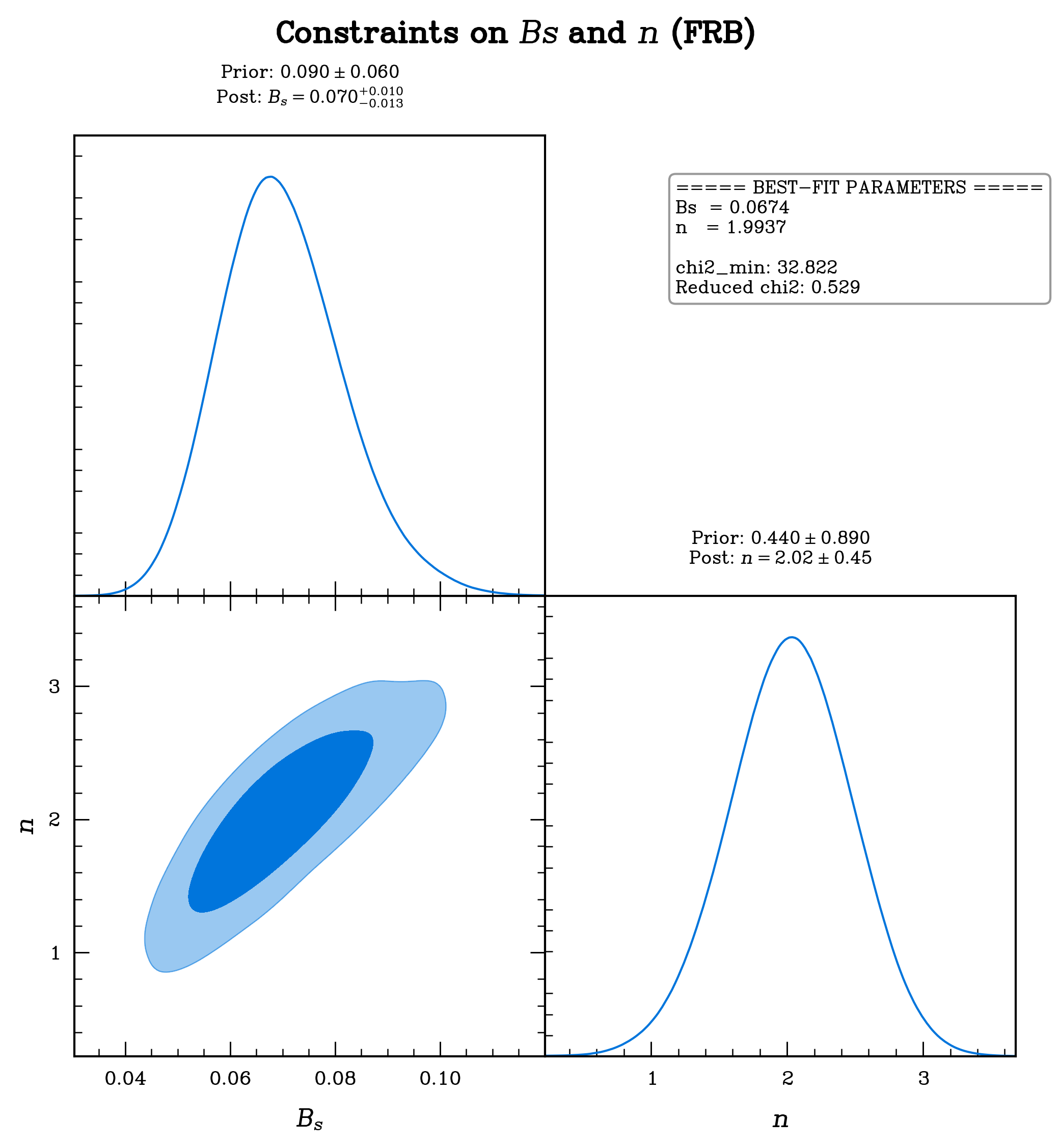}}
\hfill
\subfloat[Gaussian priors from $H(z)$]{%
\includegraphics[width=0.33\textwidth]{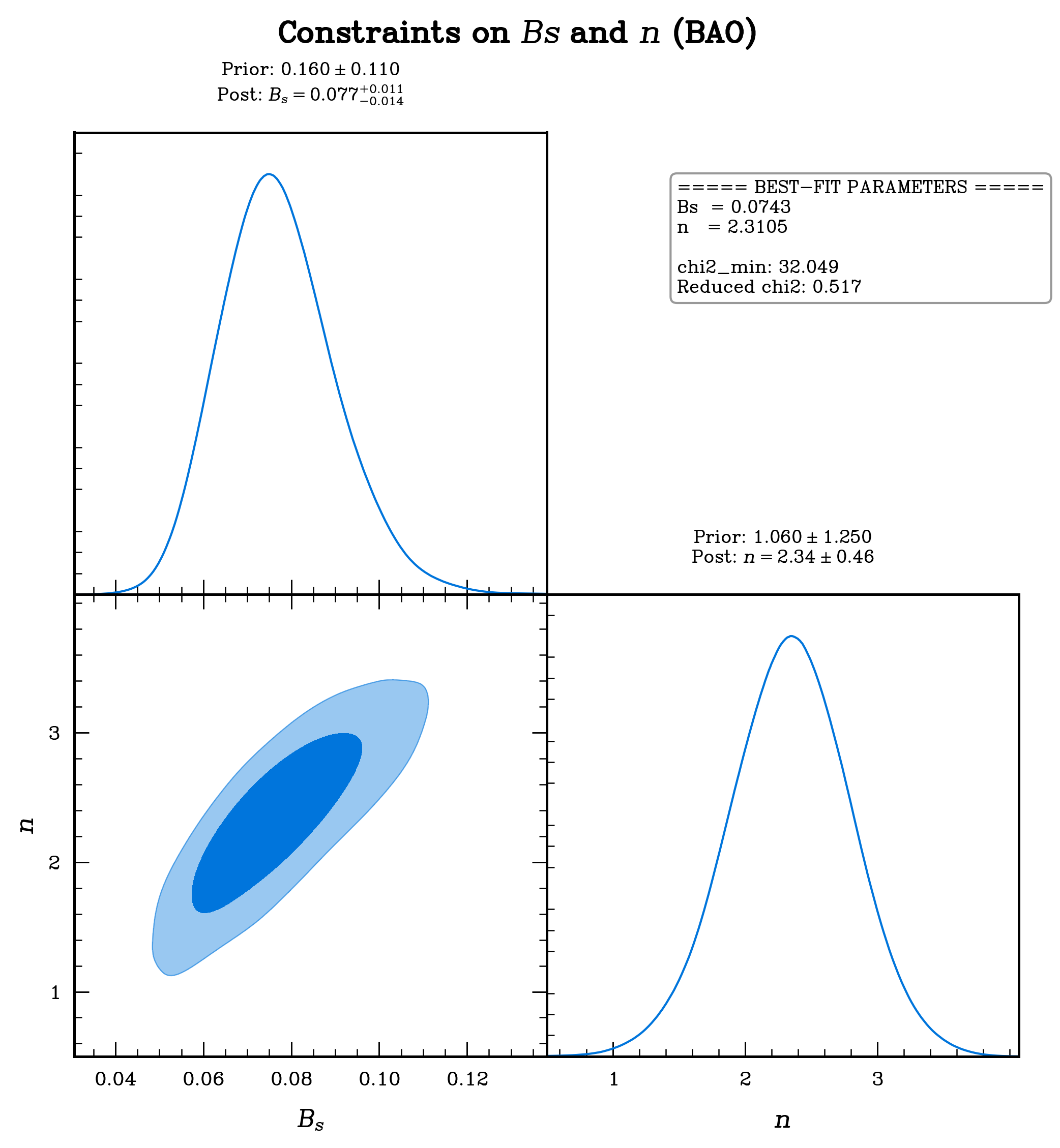}}

\vspace{0.25cm}

\subfloat[Gaussian priors from FRB]{%
\includegraphics[width=0.33\textwidth]{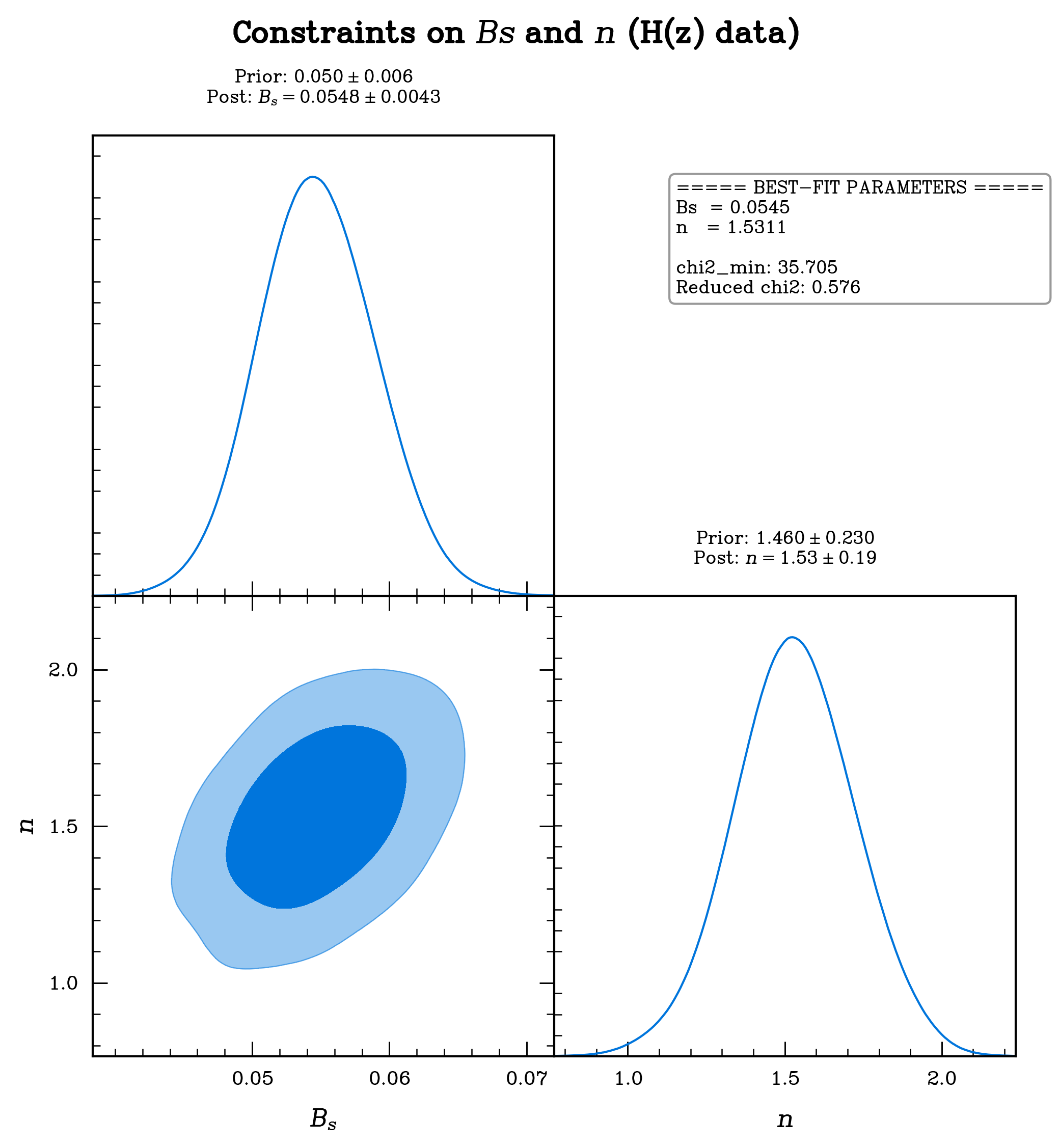}}
\hspace{0.035\textwidth} 
\subfloat[Gaussian priors from GRB]{%
\includegraphics[width=0.33\textwidth]{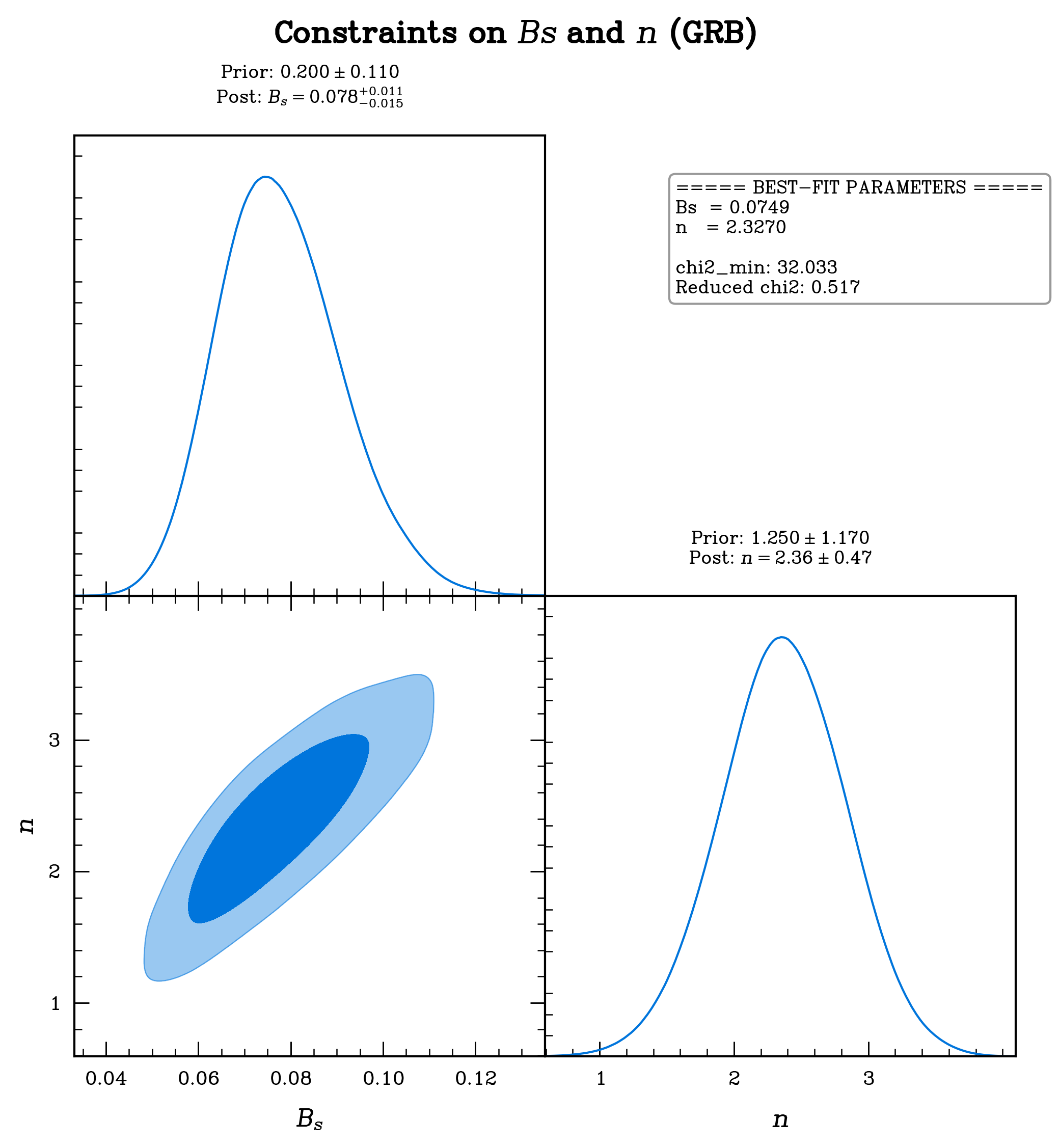}}

\caption{
Posterior distributions of the Variable Chaplygin Gas parameters $(B_s,n)$ obtained from the growth observable $f\sigma_8$. Each panel corresponds to an independent MCMC analysis employing Gaussian priors derived from previous background-only constraints based on the Pantheon, BAO, $H(z)$, FRB, and GRB datasets. The contours denote the 68\% and 95\% confidence regions.
}
\label{fig:cornerplots}
\end{figure*}

\begin{figure*}[t]
\centering
\includegraphics[width=\textwidth]{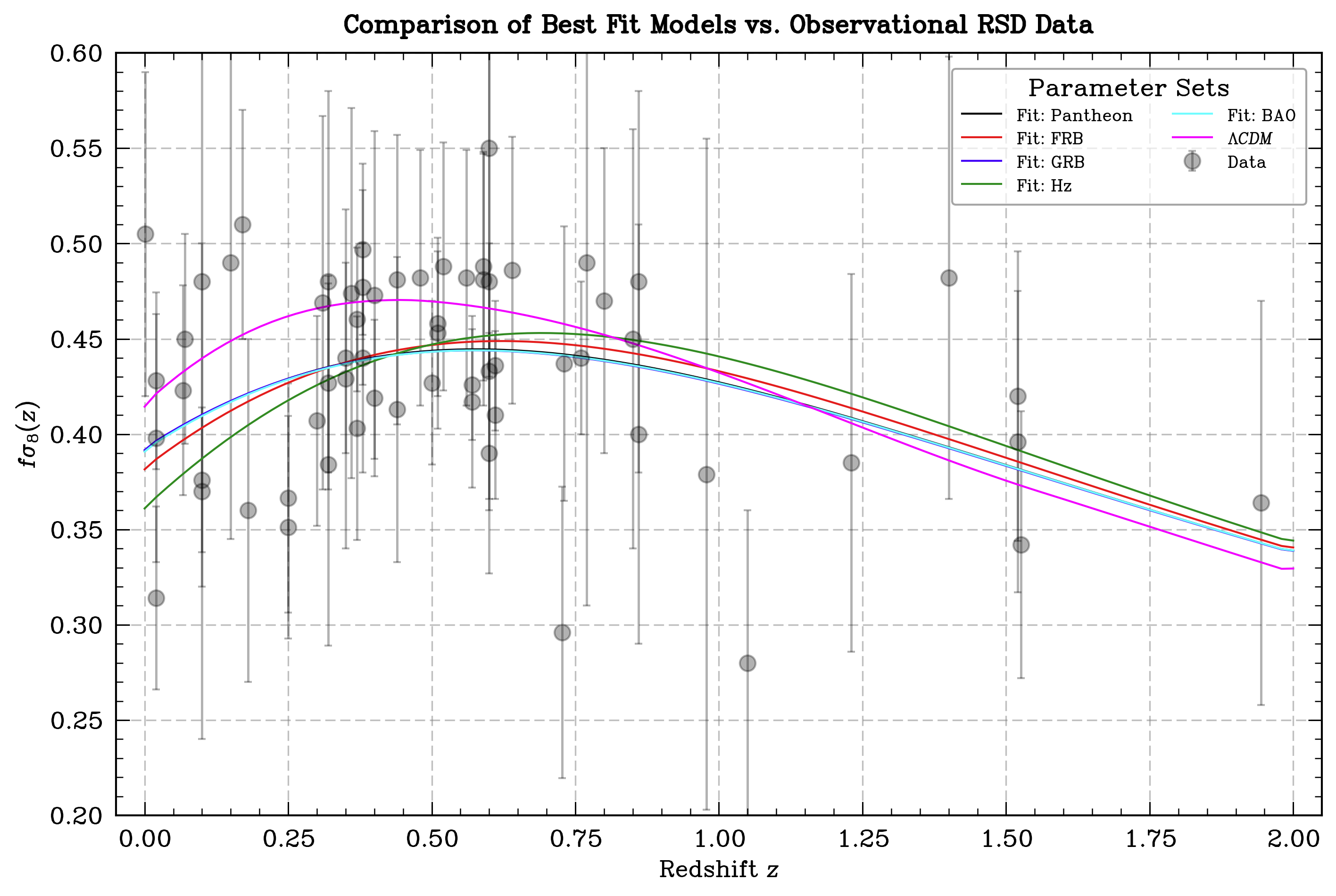}
\caption{All five VCG models are found to be consistent with the current observations within the observational uncertainties.
}
\label{fig:cornerplots1}
\end{figure*}

We constrain the parameters of the Variable Chaplygin Gas (VCG) model using growth of structure data through the observable $f\sigma_8$, and compare the results with constraints obtained from background cosmological observations. The parameter space $(B_{s},n)$ is explored using a Markov Chain Monte Carlo (MCMC) approach. The resulting marginalized posterior distributions and confidence contours are displayed in Fig. \ref{fig:cornerplots}, while the corresponding best-fit cosmological growth curves are compared against the observational data in Fig. \ref{fig:cornerplots1}. The cosmological constraints on the VCG parameters obtained from our MCMC analysis are summarized in Table~\ref{tab:mcmc_results}.

\begin{table}[h!]
    \centering
    \caption{Best-fit values and $1\sigma$ uncertainties of the VCG model parameters, $(B_s,n)$, obtained from the MCMC analysis of the $f\sigma_8$ growth data. For each analysis, Gaussian priors are adopted from the corresponding background constraints listed in Table~\ref{tab:background_constraints}.}
    \begin{tabular*}{\columnwidth}{@{\extracolsep{\fill}}lcc}
        \hline
        \hline
        \textbf{Dataset} & \bm{$B_s$} & \bm{$n$} \\
        \hline
        SNe Ia (Pantheon) & $0.077 \pm 0.015$ & $2.32 \pm 0.47$ \\
        FRB & $0.07 \pm 0.013$ & $2.02 \pm 0.45$ \\
        BAO & $0.077 \pm 0.014$ & $2.34 \pm 0.46$ \\
        $H(z)$ data & $0.058 \pm 0.004$ & $1.53 \pm 0.19$ \\
        GRB & $0.78 \pm 0.015$ & $2.36 \pm 0.47$ \\
        \hline
        \hline
    \end{tabular*}
    \label{tab:mcmc_results}
\end{table}

The analysis based on $f\sigma_8$ data yields tightly constrained parameter values:
\begin{align}
B_s &\approx 0.077 \pm 0.01, \\
n &\approx 2.3 \pm 0.1, 
\end{align}
A significant difference is observed when these constraints are compared with those obtained from background datasets. While background probes such as SNe Ia, BAO, $H(z)$, FRB, and GRB allow a broad range of parameter values, particularly for $n$, the growth data impose much tighter constraints. In particular, the value of $n$ preferred by growth observations ($n \sim 2.3$) is substantially higher than that obtained from background data ($n \sim 1$), and the parameter $B_s$ is correspondingly much smaller in the growth analysis ($\sim 0.077$) compared to the range $0.05$--$0.20$ inferred from background observations. 

This discrepancy arises due to the fundamentally different sensitivities of the datasets. Background observations constrain the expansion history of the Universe through $H(z)$, whereas growth measurements probe the evolution of matter perturbations, which depend not only on the expansion rate but also on the clustering properties of the cosmic fluid. Consequently, models that reproduce the observed expansion history may not necessarily provide a consistent description of structure formation.

The tension observed in the parameters $(B_s, n)$ suggests that the model exhibits different behaviour at the perturbation level compared to the background level. This may point to limitations of the VCG model in simultaneously explaining both expansion and growth of structure.

The results highlight the importance of incorporating growth observables such as $f\sigma_8$, which provide a bias-independent and complementary probe of cosmology. The strong constraints obtained from growth data demonstrate that perturbation-based analyses are essential for breaking degeneracies present in background-only studies.

In conclusion, while the Variable Chaplygin Gas model remains a viable candidate for describing the dark sector at the background level, our analysis shows that growth of structure data place stringent constraints that reveal a significant tension with background results. A consistent cosmological model must therefore satisfy both expansion and perturbation constraints simultaneously. Future high-precision surveys such as DESI and Euclid are expected to further refine these constraints and provide deeper insights into the viability of unified dark energy models.
\section{Acknowledgement }
The authors thank Prof. T. R. Seshadri and Dr. Sanil Unnikrishnan for useful discussions and comments. RJ thanks GS and ST for the opportunity to work on this project, and Dr. Swagat Saurav Mishra for his guidance, support, and useful discussions. The authors are also grateful to St. Stephen's College for providing support.
\bibliographystyle{apsrev4-2}
\bibliography{references}

@article{Copeland2006,
  author = {Copeland, E. J. and Sami, M. and Tsujikawa, S.},
  title = {Dynamics of Dark Energy},
  journal = {International Journal of Modern Physics D},
  volume = {15},
  pages = {1753--1936},
  year = {2006}
}

@article{Padmanabhan2003,
  author = {Padmanabhan, T.},
  title = {Cosmological Constant—the Weight of the Vacuum},
  journal = {Physics Reports},
  volume = {380},
  pages = {235--320},
  year = {2003}
}

@article{Kamenshchik2001,
  author = {Kamenshchik, A. Y. and Moschella, U. and Pasquier, V.},
  title = {An Alternative to Quintessence},
  journal = {Physics Letters B},
  volume = {511},
  pages = {265--268},
  year = {2001}
}

@article{Bento2002,
  author = {Bento, M. C. and Bertolami, O. and Sen, A. A.},
  title = {Generalized Chaplygin Gas, Accelerated Expansion and Dark-Energy-Matter Unification},
  journal = {Physical Review D},
  volume = {66},
  pages = {043507},
  year = {2002}
}

@article{Guo2007,
  author = {Guo, Z. K. and Zhang, Y. Z.},
  title = {Cosmological Evolution of Variable Chaplygin Gas},
  journal = {Physics Letters B},
  volume = {645},
  pages = {326--329},
  year = {2007}
}

@article{Linder2005,
  author = {Linder, Eric V.},
  title = {Cosmic Growth History and Expansion History},
  journal = {Physical Review D},
  volume = {72},
  pages = {043529},
  year = {2005}
}

@article{Nesseris2017,
  author = {Nesseris, S. and Basilakos, S. and Saridakis, E. N. and Perivolaropoulos, L.},
  title = {Viable f(R) Gravity Models and Their Confrontation with Growth Data},
  journal = {Physical Review D},
  volume = {96},
  pages = {023542},
  year = {2017}
}

@article{DESI2024,
  author = {{DESI Collaboration}},
  title = {DESI 2024 Results: Cosmological Constraints from the First Year},
  journal = {arXiv e-prints},
  eprint = {2404.03000},
  archivePrefix = {arXiv},
  primaryClass = {astro-ph.CO},
  year = {2024}
}

@article{Clifton2012,
  author = {Clifton, T. and Ferreira, P. G. and Padilla, A. and Skordis, C.},
  title = {Modified Gravity and Cosmology},
  journal = {Physics Reports},
  volume = {513},
  pages = {1--189},
  year = {2012}
}

@article{Tsujikawa2008,
  author = {Tsujikawa, S.},
  title = {Observational Signatures of f(R) Dark Energy Models},
  journal = {Physical Review D},
  volume = {77},
  pages = {023507},
  year = {2008}
}

@article{Thakur2011,
  author  = {Shruti Thakur and Anjan A. Sen and T. R. Seshadri},
  title   = {Non-minimally coupled f(R) cosmology},
  journal = {Physics Letters B},
  volume  = {696},
  number  = {4},
  pages   = {309--314},
  year    = {2011},
  doi     = {10.1016/j.physletb.2010.12.043},
  eprint  = {1007.5250},
  archivePrefix = {arXiv},
  primaryClass  = {astro-ph.CO}
}

@article{Thakur2013,
  author  = {Shruti Thakur and Anjan A. Sen},
  title   = {Can structure formation distinguish {$\Lambda$}CDM from non-minimal f(R) gravity?},
  journal = {Physical Review D},
  volume  = {88},
  number  = {4},
  pages   = {044043},
  year    = {2013},
  doi     = {10.1103/PhysRevD.88.044043},
  eprint  = {1305.0420},
  archivePrefix = {arXiv},
  primaryClass  = {gr-qc}
}

@article{Nguyen2023,
  author  = {Nhat-Minh Nguyen and Dragan Huterer and Yuewei Wen},
  title   = {Evidence for Suppression of Structure Growth in the Concordance Cosmological Model},
  journal = {Physical Review Letters},
  volume  = {131},
  number  = {11},
  pages   = {111001},
  year    = {2023},
  doi     = {10.1103/PhysRevLett.131.111001},
  eprint  = {2302.01331},
  archivePrefix = {arXiv},
  primaryClass  = {astro-ph.CO}
}

@article{Planck2018,
  author  = {{Planck Collaboration}},
  title   = {Planck 2018 Results. VI. Cosmological Parameters},
  journal = {Astronomy \& Astrophysics},
  volume  = {641},
  pages   = {A6},
  year    = {2020},
  doi     = {10.1051/0004-6361/201833910},
  eprint  = {1807.06209},
  archivePrefix = {arXiv},
  primaryClass  = {astro-ph.CO}
}

@article{Euclid2024,
  author  = {{Euclid Collaboration}},
  title   = {Euclid Preparation: VII. Forecast Validation for Euclid Cosmological Probes},
  journal = {Astronomy \& Astrophysics},
  volume  = {642},
  pages   = {A191},
  year    = {2020},
  doi     = {10.1051/0004-6361/202038071}
}

@article{Sethi2024,
  author  = {Geetanjali Sethi and Udish Sharma and Nadia Makhijani},
  title   = {Variable Chaplygin Gas: Constraining Parameters Using Fast Radio Bursts},
  journal = {Astrophysics and Space Science},
  volume  = {369},
  pages   = {42},
  year    = {2024},
  doi     = {10.1007/s10509-024-04306-6}
}

@article{Sen2005,
  author       = {Anjan A. Sen and Robert J. Scherrer},
  title        = {Generalizing the Generalized Chaplygin Gas},
  journal      = {Physical Review D},
  volume       = {72},
  number       = {6},
  pages        = {063511},
  year         = {2005},
  doi          = {10.1103/PhysRevD.72.063511},
  eprint       = {astro-ph/0507717},
  archivePrefix= {arXiv},
  primaryClass = {astro-ph}
}

@article{Bhardwaj2024,
  author  = {Yogesh Bhardwaj and C. P. Singh},
  title   = {Constraining the Variable Generalized Chaplygin Gas Model in Matter Creation Cosmology},
  journal = {Communications in Theoretical Physics},
  volume  = {76},
  number  = {10},
  pages   = {105403},
  year    = {2024},
  doi     = {10.1088/1572-9494/ad58c2}
}

@article{Mukherjee2024,
  author  = {Puja Mukherjee and Ujjal Debnath and Himanshu Chaudhary and G. Mustafa},
  title   = {Constraining the Parameters of Generalized and Viscous Modified Chaplygin Gas and Black Hole Accretion in Einstein--Aether Gravity},
  journal = {European Physical Journal C},
  volume  = {84},
  pages   = {930},
  year    = {2024},
  doi     = {10.1140/epjc/s10052-024-13196-5}
}

@article{Bhardwaj2025,
  author  = {Yogesh Bhardwaj and C. P. Singh},
  title   = {Cosmological Dynamics of Matter Creation with Modified Chaplygin Gas and Bulk Viscosity},
  journal = {European Physical Journal C},
  volume  = {85},
  pages   = {1465},
  year    = {2025},
  doi     = {10.1140/epjc/s10052-025-15227-1}
}

@article{Aggarwal2025,
  author  = {Nakul Aggarwal and Ali Pourmand and Fatimah Shojai and Harish Parthasarathy},
  title   = {Observational Constraints on Chaplygin Gas Models in Non-Minimally Coupled Power-Law $f(Q)$ Gravity with Quasars},
  journal = {arXiv e-prints},
  year    = {2025},
  eprint  = {2510.12472},
  archivePrefix = {arXiv},
  primaryClass  = {astro-ph.CO}
}

@article{AlMamon2026,
  author  = {Abdulla Al Mamon and Andronikos Paliathanasis and Subhajit Saha},
  title   = {Unifying the Dark Sector with the New Generalized Chaplygin Gas: Observational Constraints},
  journal = {Fortschritte der Physik},
  volume  = {74},
  number  = {6},
  pages   = {e70127},
  year    = {2026},
  doi     = {10.1002/prop.70127}
}

@article{Sandvik2004,
  author  = {H{\aa}vard B. Sandvik and Max Tegmark and Matias Zaldarriaga and Ioav Waga},
  title   = {The End of Unified Dark Matter?},
  journal = {Physical Review D},
  volume  = {69},
  pages   = {123524},
  year    = {2004},
  doi     = {10.1103/PhysRevD.69.123524},
  eprint  = {astro-ph/0212114},
  archivePrefix = {arXiv}
}

@article{Bean2003,
  author  = {Rachel Bean and Olivier Dor{\'e}},
  title   = {Are Chaplygin Gas Cosmologies Viable?},
  journal = {Physical Review D},
  volume  = {68},
  pages   = {023515},
  year    = {2003},
  doi     = {10.1103/PhysRevD.68.023515},
  eprint  = {astro-ph/0301308},
  archivePrefix = {arXiv}
}

@article{Fabris2010,
  author  = {Julio C. Fabris and H. E. S. Velten and Winfried Zimdahl},
  title   = {Matter Power Spectrum for the Generalized Chaplygin Gas Model: The Relativistic Case},
  journal = {Physical Review D},
  volume  = {81},
  pages   = {087303},
  year    = {2010},
  doi     = {10.1103/PhysRevD.81.087303},
  eprint  = {1001.4101},
  archivePrefix = {arXiv}
}

@article{DelPopolo2013,
  author       = {A. Del Popolo and F. Pace and S. P. Maydanyuk and J. A. S. Lima and J. F. Jesus},
  title        = {Shear and Rotation in Chaplygin Cosmology},
  journal      = {Physical Review D},
  volume       = {87},
  number       = {4},
  pages        = {043527},
  year         = {2013},
  doi          = {10.1103/PhysRevD.87.043527},
  eprint       = {1303.3628},
  archivePrefix= {arXiv},
  primaryClass = {astro-ph.CO}
}

@article{BBKS1986,
  author  = {J. M. Bardeen and J. R. Bond and N. Kaiser and A. S. Szalay},
  title   = {The Statistics of Peaks of Gaussian Random Fields},
  journal = {Astrophysical Journal},
  volume  = {304},
  pages   = {15--61},
  year    = {1986},
  doi     = {10.1086/164143}
}

@article{Eisenstein1998,
  author  = {Daniel J. Eisenstein and Wayne Hu},
  title   = {Baryonic Features in the Matter Transfer Function},
  journal = {Astrophysical Journal},
  volume  = {496},
  pages   = {605--614},
  year    = {1998},
  doi     = {10.1086/305424},
  eprint  = {astro-ph/9709112},
  archivePrefix = {arXiv}
}

@article{Song2009,
  author        = {Song, Yong-Seon and Percival, Will J.},
  title         = {Reconstructing the History of Structure Formation using Redshift Distortions},
  journal       = {Journal of Cosmology and Astroparticle Physics},
  year          = {2009},
  volume        = {10},
  number        = {10},
  pages         = {004},
  doi           = {10.1088/1475-7516/2009/10/004},
  eprint        = {0807.0810},
  archivePrefix = {arXiv},
  primaryClass  = {astro-ph},
  url           = {https://arxiv.org/abs/0807.0810}
}

@article{Beutler2012,
  author  = {Florian Beutler et al.},
  title   = {The 6dF Galaxy Survey: z≈0 Measurements of the Growth Rate and $\sigma_8$},
  journal = {Monthly Notices of the Royal Astronomical Society},
  volume  = {423},
  pages   = {3430--3444},
  year    = {2012},
  doi     = {10.1111/j.1365-2966.2012.21136.x},
  eprint  = {1204.4725},
  archivePrefix = {arXiv}
}

@article{Huterer2023,
  author  = {Dragan Huterer},
  title   = {Growth of Cosmic Structure},
  journal = {The Astronomy and Astrophysics Review},
  volume  = {31},
  number  = {1},
  pages   = {2},
  year    = {2023},
  doi     = {10.1007/s00159-023-00147-4},
  eprint  = {2212.05003},
  archivePrefix = {arXiv},
  primaryClass  = {astro-ph.CO}
}

@article{Kazantzidis2021,
  author  = {L. Kazantzidis and L. Perivolaropoulos},
  title   = {Evolution of the $f\sigma_8$ Tension with the Planck15/$\Lambda$CDM Determination and Implications for Modified Gravity Theories},
  journal = {Phys. Rev. D},
  volume  = {105},
  pages   = {023520},
  year    = {2022},
  doi     = {10.1103/PhysRevD.105.023520}
}

@article{Kazantzidis2018,
  author       = {L. Kazantzidis and L. Perivolaropoulos},
  title        = {Evolution of the {$f\sigma_8$} Tension with the Planck15/$\Lambda$CDM Determination and Implications for Modified Gravity Theories},
  journal      = {Physical Review D},
  volume       = {97},
  number       = {10},
  pages        = {103503},
  year         = {2018},
  doi          = {10.1103/PhysRevD.97.103503},
  eprint       = {1803.01337},
  archivePrefix= {arXiv},
  primaryClass = {astro-ph.CO}
}

@article{BouhmadiLopez2025,
  author       = {Mariam Bouhmadi-L{\'o}pez and Be{\~n}at Ibarra-Uriondo},
  title        = {Cosmological Perturbations for Smooth Sign-Switching Dark Energy Models},
  journal      = {Physics of the Dark Universe},
  volume       = {50},
  pages        = {102129},
  year         = {2025},
  doi          = {10.1016/j.dark.2025.102129},
  eprint       = {2506.18992},
  archivePrefix= {arXiv},
  primaryClass = {gr-qc}
}

@article{lai2026desi,
  title={The DESI DR1 Peculiar Velocity Survey: growth rate measurements from the maximum likelihood fields method},
  author={Lai, Y and Howlett, C and Aguilar, J and Ahlen, S and Amsellem, AJ and Bautista, J and BenZvi, S and Bianchi, D and Blake, C and Brooks, D and others},
  journal={Journal of Cosmology and Astroparticle Physics},
  volume={2026},
  number={04},
  pages={026},
  year={2026},
  publisher={IOP Publishing}
}

@book{Peebles1980,
  author    = {Peebles, P. J. E.},
  title     = {The Large-Scale Structure of the Universe},
  publisher = {Princeton University Press},
  address   = {Princeton, New Jersey},
  year      = {1980},
  isbn      = {9780691082400},
  adsurl    = {https://ui.adsabs.harvard.edu/abs/1980lssu.book.....P}
}

@article{Sethi2006,
    author = {Sethi, Geetanjali and Singh, Sushil K. and Kumar, Pranav and Jain, Deepak and Dev, Abha},
    title = {Variable Chaplygin Gas: Constraints from CMBR and SNe Ia},
    journal = {International Journal of Modern Physics D},
    volume = {15},
    number = {07},
    pages = {1089--1098},
    year = {2006},
    doi = {10.1142/S0218271806008644}
}
\end{document}